\documentclass[preprintnumbers,article,amsmath,amssymb,floatfix,10pt,prd,onecolumn,
superscriptaddress,nofootinbib]{revtex4-2}
\usepackage{bm}
\usepackage{amsfonts}
\usepackage{latexsym}
\usepackage[latin1]{inputenc}
\usepackage{graphicx}
\usepackage{amsmath}
\usepackage{palatino}
\usepackage{mathpazo}
\usepackage{textcomp}
\usepackage{float}
\usepackage{booktabs}
\usepackage{dcolumn}
\usepackage{ragged2e}
\usepackage{hyperref}
\usepackage{tensor}
\usepackage[version=3]{mhchem}%to write chemical symbols
\hypersetup{colorlinks,citecolor=blue}
\hypersetup{colorlinks=true,linkcolor=red,filecolor=magenta,    urlcolor=blue}
\usepackage{amsmath}
\usepackage{xcolor}
\usepackage{enumitem}
\usepackage{orcidlink}
\usepackage{epsfig}
\usepackage{subfigure}
\usepackage{commath}
\usepackage{caption}
\def\jnl@style{\it}
\def\aaref@jnl#1{{\jnl@style#1}}

\def\aaref@jnl#1{{\jnl@style#1}}

\def\aj{\aaref@jnl{AJ}}                   % Astronomical Journal
\def\apj{\aaref@jnl{ApJ}}                 % Astrophysical Journal
\def\apjl{\aaref@jnl{ApJ}}                % Astrophysical Journal, Letters
\def\apjs{\aaref@jnl{ApJS}}               % Astrophysical Journal, Supplement
\def\apss{\aaref@jnl{Ap\&SS}}             % Astrophysics and Space Science
\def\aap{\aaref@jnl{A\&A}}                % Astronomy and Astrophysics
\def\aapr{\aaref@jnl{A\&A~Rev.}}          % Astronomy and Astrophysics Reviews
\def\aaps{\aaref@jnl{A\&AS}}              % Astronomy and Astrophysics, Supplement
\def\mnras{\aaref@jnl{Mon.~Not.~Roy.~Astron.~Soc.}}             % Monthly Notices of the RAS
\def\prd{\aaref@jnl{Phys.~Rev.~D}}        % Physical Review D
\def\prc{\aaref@jnl{Phys.~Rev.~C}}  % Physical Review C
\def\prl{\aaref@jnl{Phys.~Rev.~Lett.}}    % Physical Review Letters
\def\qjras{\aaref@jnl{QJRAS}}             % Quarterly Journal of the RAS
\def\skytel{\aaref@jnl{S\&T}}             % Sky and Telescope
\def\ssr{\aaref@jnl{Space~Sci.~Rev.}}     % Space Science Reviews
\def\zap{\aaref@jnl{ZAp}}                 % Zeitschrift fuer Astrophysik
\def\nat{\aaref@jnl{Nature}}              % Nature
\def\aplett{\aaref@jnl{Astrophys.~Lett.}} % Astrophysics Letters
\def\apspr{\aaref@jnl{Astrophys.~Space~Phys.~Res.}} % Astrophysics Space Physics Research
\def\physrep{\aaref@jnl{Phys.~Rep.}}      % Physics Reports
\def\physscr{\aaref@jnl{Phys.~Scr}}       % Physica Scripta
\def\commat{\aaref@jnl{Comm.~Math.~Phys.}}              % Communications in Mathematical Physics
\def\science{\aaref@jnl{Science}}               % Science
\def\cqg{\aaref@jnl{Classical Quant.~Grav.}}            % Classical and Quantum Gravity
\def\jpcs{\aaref@jnl{JPCS}}                                     % Journal of Physics Conference Series
\def\ijmpd{\aaref@jnl{Int.~J.~Mod.~Phys.~D}}                    % International Journal of Modern Physics D
\def\grg{\aaref@jnl{Gen.~Relat.~Gravit.}}               % General Relativity and Gravitation
\def\rpp{\aaref@jnl{Rep.~Prog.~Phys.}}          % Reports on Progress in Physics
\def\npa{\aaref@jnl{Nucl.~Phys.~A}}        % Nuclear Physics A
\def\lrr{\aaref@jnl{Living Rev.~Rel.}}                   % Living reviews in relativity
\def\jcap{\aaref@jnl{J.~Cosmology Astropart.~Phys.}}    % Journal of cosmology and astroparticle physics
\def\rmp{\aaref@jnl{Rev.~Mod.~Phys.}}   %Reviews of modern physics
\def\epjc{\aaref@jnl{Eur.~Phys.~J.~C}} 
\def\plb{\aaref@jnl{~Phy.~Lett.~B}} 
\def\mpla{\aaref@jnl{Mod.~Phy.~Lett.~A}} 
\def\arxiv{\aaref@jnl{arxiv.org}}

\allowdisplaybreaks[1]
\date{\today}

\begin{document}

\title{Chebyshev cosmography in the framework of extended symmetric teleparallel theory}

\author{Sai Swagat Mishra\orcidlink{0000-0003-0580-0798}}
\email{saiswagat009@gmail.com}
\affiliation{Department of Mathematics, Birla Institute of Technology and
Science-Pilani,\\ Hyderabad Campus, Hyderabad-500078, India.}

\author{N. S. Kavya\orcidlink{0000-0001-8561-130X}}
\email{kavya.samak.10@gmail.com}
\affiliation{Department of Mathematics, Birla Institute of Technology and
Science-Pilani,\\ Hyderabad Campus, Hyderabad-500078, India.}
\affiliation{Department of P.G. Studies and Research in Mathematics,
 \\
 Kuvempu University, Shankaraghatta, Shivamogga 577451, Karnataka, India.
}%

\author{P.K. Sahoo\orcidlink{0000-0003-2130-8832}}
\email{pksahoo@hyderabad.bits-pilani.ac.in}
\affiliation{Department of Mathematics, Birla Institute of Technology and
Science-Pilani,\\ Hyderabad Campus, Hyderabad-500078, India.}

\author{V. Venkatesha\orcidlink{0000-0002-2799-2535}}%
 \email{vensmath@gmail.com}
\affiliation{Department of P.G. Studies and Research in Mathematics,
 \\
 Kuvempu University, Shankaraghatta, Shivamogga 577451, Karnataka, India.
}%

\begin{abstract}
   Cosmography has been extensively utilized to constrain the kinematic state of the Universe using measured distances. % Being a model-independent approach it relies solely on the assumption that the Universe is homogeneous and isotropic. 
   In this work, we propose a new method to reconstruct coupling theories using the first kind of Chebyshev polynomial for two variables in which the functional form of the $f(Q,T)$ theory has been obtained. Further, the unknowns that appeared in the series are constrained using the cosmographic parameters. We find the explicit form of the luminosity distance in terms of cosmographic parameters to perform MCMC analysis using the PANTHEON+SH0ES data set. Through the distance modulus function, we observe that the result comes out to be an excellent match to the standard cosmological model and data.

\end{abstract}
\maketitle
\section{Introduction}

The standard model of cosmology, known as the $\Lambda$CDM paradigm, presumes general relativity (GR) as the fundamental gravitational theory and incorporates the standard model of particles, cold dark matter (CDM), and the cosmological constant $\Lambda$. This model has been extensively validated by various observational datasets, including Type Ia supernovae \cite{SupernovaSearchTeam:1998fmf,SupernovaCosmologyProject:1998vns,SupernovaSearchTeam:2004lze}, baryon acoustic oscillations \cite{SDSS:2005xqv,SDSS:2009ocz}, the cosmic microwave background \cite{WMAP:2003elm}, and weak lensing experiments \cite{Heymans:2012gg,Shi:2017qpr}.

However, recent observational findings have uncovered potential tensions, such as the Hubble tension and the $\sigma_8$ tension between early-time measurements under $\Lambda$CDM and late-time probes \cite{WMAP:2006bqn,WMAP:2010qai,Demianski:2016zxi}. Additionally, the non-renormalizability of GR and its challenges in achieving a quantum description present significant drawbacks for the theory. As a result, considerable efforts have been directed towards developing various modifications to the theory of gravity, with the goal of addressing or resolving these issues \cite{Joyce:2014kja,Tsujikawa:2003jp,Velten:2014nra,Astesiano:2022ozl,Katsuragawa:2016yir,Zaregonbadi:2016xna,Joudaki:2016kym,Mandal:2023bzo,Kavya:2024ssu}. 

The evolution of gravitational theories has paralleled the progress in differential geometry. According to the geometric frameworks of gravity, it is posited that spacetime is equipped with a metric structure within a general space. This structure is defined by the line element  $ds = F(x^1, \ldots, x^n; dx^1, \ldots, dx^n)$, where  $F(x; \xi)>0$ for  $\xi \neq 0$ defined on the tangent bundle $T_M$ with $F$ being homogeneous of degree 1 in $\xi$. Particularly, $F^2 = g_{\mu\nu}dx^{\mu}dx^{\nu}$ corresponds to the Riemannian geometry. One possible way to modify GR is through the modification of gravitational Lagrangian description while keeping the same geometric framework. Consequently, theories such as $f(R)$, $f(R,T)$, $f(R,\mathcal{L}_m)$ have emerged \cite{DeFelice:2010aj,Harko:2011kv,Moraes:2017mir,Kavya:2023tjf}. These can be considered extensions of GR, with GR being the limiting case. Another interesting aspect of modifying GR involves considering a more generalized geometric explanation, initially introduced by Weyl and Cartan \cite{Weyl:1918ib,Trautman:2006fp}. This gave rise to gravitational theories such as Riemann-Cartan and Weyl-Cartan. It was further developed by the notion of changing the standard connection to a general affine connection, which includes torsion and non-metricity contributions \cite{Capozziello:2022zzh}. 

In the present work, we focus on one of these theories, in which non-metricity acts as a key geometric element in describing spacetime. The non-metricity describes situations where the metric tensor varies in a way that is not preserved by parallel transport. Mathematically, $\tilde{\nabla}_{\alpha}g_{\mu\nu}=0$, where $\tilde{\nabla}_{\alpha}$ is the covarient derivative with respect to Levi-Civita connection. This approach helps us explain the GR equivalent scenario in terms of an entirely curvatureless and torsionless geometric background. Although fundamentally, the Symmetric Teleparallel equivalent of GR produces similar outcomes to GR, the corresponding extensions do not necessarily yield the same results. Motivated by this, in this article, we work on non-metricity-based theory. As a simple extension, one can consider $f(Q)$ gravity, where the gravitational Lagrangian is a functional form of non-metricity scalar \cite{BeltranJimenez:2019tme,Mishra:2024rut,Kavya:2024bpj}. Recently, Xu et al. \cite{Xu:2019sbp} proposed a well-known coupling of $f(Q)$ gravity called  $f(Q,T)$ theory, in which the matter sector is coupled with the geometric sector. In the literature, a plethora of works have been carried out on the astronomical and cosmological implications of this modified gravity \cite{Kavya:2024kdi, Bhattacharjee:2020wbh, Gadbail:2023klq, Xu:2020yeg, Arora:2020tuk, Yang:2021fjy, Arora:2020met, Najera:2021afa}. To assess the viability of $f(Q,T)$ gravity, we adopt a model-independent approach. This is achieved through the cosmographic reconstruction technique. 

Cosmography is a descriptive framework that focuses on the large-scale structure and observable features of the Universe, without necessarily going deep into the underlying physical laws or models that govern its behavior. It serves as a way to map and describe the Universe's characteristics as observed, providing a bridge between observational data and theoretical cosmology. It does not rely on any prior assumptions of a cosmological model. This technique utilizes a set of cosmographic parameters, which are constrained using observational data to analyze the dynamics of the Universe. In the literature, several works have been carried out considering the Taylor series expansion, which works well in the vicinity of late-time redshifts (particularly  $z < 1$) \cite{Capozziello:2014zda, Luongo:2012dv, Mishra:2024oln,Capozziello:2008qc,Capozziello:2011hj,Sabiee:2022iyo,Mandal:2020buf, Gao:2021mnras, Farias:2021jdz}. Recently, an alternative series expansion using Chebyshev polynomials has been studied \cite{Capozziello:2017nbu}, which may overcome this issue. In this work, we propose a novel way of using multivariable Chebyshev polynomials to measure coupled $f(Q,T)$  coefficients. 

The layout of the article is as follows: the fundamentals and geometric foundations of $f(Q,T)$  cosmology are discussed in \autoref{sec:II}. In \autoref{sec:III}, we present the standard cosmographic assumptions of the series form in terms of cosmographic parameters. The main highlight of our work, which describes the functions using Chebyshev polynomials, is presented in \autoref{sec:IV}. The parameters are constrained using empirical data. The data and methodology used are given in \autoref{sec:V}. Finally, in \autoref{sec:VI}, we conclude and summarize our results.

\section{Fundamentals of $f(Q,T)$ theory}\label{sec:II}
An initial modification to GR involves generalizing the definition of affine connections, incorporating components beyond Levi-Civita. This leads to the formalism of a metric-affine theory, characterized by the triplet $\{M, g_{\mu\nu}, \Gamma^{\rho}_{\mu\nu}\}$. Here, $M$ represents a four-dimensional spacetime manifold, $g_{\mu\nu}$ denotes a rank-two symmetric tensor with 10 independent components, and $\Gamma^{\rho}_{\mu\nu}$ signifies the affine connection which can be uniquely decomposed as \cite{BeltranJimenez:2019esp}

\begin{equation}
    \Gamma^{\rho}_{\mu\nu} = \tilde{\Gamma}^{\rho}_{\mu\nu} + K^{\rho}_{\mu\nu} + L^{\rho}_{\mu\nu},
\end{equation}
where $\tilde{\Gamma}^{\rho}_{\mu\nu}:= \frac{1}{2} g^{\rho\lambda} (\partial_{\mu} g_{\lambda\nu} + \partial_{\nu} g_{\mu\lambda} - \partial_{\lambda} g_{\mu\nu}) $ is the Levi-Civita connection, $K^{\rho}_{\mu\nu}:=\frac{1}{2} (T^{\rho}_{\mu \nu} + T^{\rho}_{\nu \mu} - T^{\rho}_{\mu\nu})$ is the contortion tensor, and $L^{\rho}_{\mu\nu}:=\frac{1}{2} (Q^{\rho}_{\mu\nu} - Q^{\rho}_{\mu \nu} - Q^{\rho}_{\nu \mu})$ is the disformation tensor. The corresponding geometric entities are represented by 
the curvature tensor $R^\mu_{\nu \alpha \beta}:= \partial_\rho \Gamma^\mu_{\nu \sigma} - \partial_\sigma \Gamma^\mu_{\nu \rho} + \Gamma^\mu_{\tau \rho} \Gamma^\tau_{\nu \sigma} - \Gamma^\mu_{\tau \sigma} \Gamma^\tau_{\nu \rho}$, the torsion tensor  $T^\mu_{\nu \rho}:= 2 \Gamma^\mu_{[\rho \nu]} \equiv \Gamma^\mu_{\rho \nu} - \Gamma^\mu_{\nu \rho}$, and the non-metricity tensor  $Q_{\rho \mu \nu} := \nabla_\mu g_{\nu \rho} \equiv \partial_\mu g_{\nu \rho} - \Gamma^\lambda_{(\nu | \mu} g_{\rho) \lambda} \neq 0$. In general, for nonmetricity-based theories, the first two geometric objects (curvature tensor and torsion tensor) vanish. %The disformation tensor can be written as 
%\begin{equation}
%    {L^\mu}_{\rho\beta} \equiv -\frac{1}{2} g^{\mu\nu} \left( \nabla_{\rho} g_{\nu\beta} + \nabla_{\beta} g_{\nu\rho} - \nabla_{\nu} g_{\rho\beta} \right).
%\end{equation}

%Using this expression, the non-metricity scalar $Q$ can be written as

Using the expression of disformation tensor, the non-metricity scalar $Q$ can be written as

\begin{equation}
    Q \equiv -g^{\rho\beta} \left( {L^{\mu}}_{\nu\rho} {L^{\nu}}_{\beta\mu} - {L^{\mu}}_{\nu\mu} {L^{\nu}}_{\rho\beta} \right).
\end{equation}

Additionally, $\nabla_{\lambda} \overset{\circ}{=} \partial_{\lambda}$ implies $Q \overset{\circ}{=} -\mathcal{L}_E$, where `$\circ$' denotes gauge coincidence and 
\begin{equation}
    \mathcal{L}_E=g^{\alpha\beta}\left(\left\{{^{\lambda}}_{\mu\alpha}\right\}\left\{{^{\mu}}_{\beta\lambda}\right\}-\left\{{^{\lambda}}_{\mu\lambda}\right\}\left\{{^{\mu}}_{\alpha\beta}\right\}\right)
\end{equation}
is the Lagrangian for the motion equations proposed by Einstein. The non-metricity tensor $Q_{\lambda\rho\beta}$ is characterized by two of its traces, which can be expressed as

\begin{equation}
    Q_\lambda = Q^{\;\;\;\rho}_{\lambda\;\;\rho}, \quad \Tilde{Q}_\lambda = {Q^{\rho}}_{\lambda\rho}.
\end{equation}

Non-metricity fails to preserve the length of vectors. To ensure length conservation, we consider the conditions $Q_{(\lambda \mu \nu)} = 0$ and $Q_{\lambda (\mu \nu)} = 0$. The existence of non-metricity introduces specific geometric effects, leading to such unique consequences compared to GR. Furthermore, the change of indices under the covariant derivative has a different description. The deviation of the anomalous acceleration indicates that the four-velocity is no longer orthogonal to the four-acceleration. However, it is possible to obtain the GR equivalent scenario known as the Symmetric Teleparallel Equivalent of General Relativity (STEGR) \cite{BeltranJimenez:2017tkd}. The action for STEGR is given by
\begin{equation}
    S_{\text{STEGR}}=\int \left(\frac{1}{2}\, \mathcal{L}_{\text{STEGR}} +\mathcal{L}_m\right)	\sqrt{-g}\, d^4x.
\end{equation}
Here, $\mathcal{L}_{\text{STEGR}}$ represents the non-metricity scalar $Q$, $g$ is the determinant of metric tensor, and $\mathcal{L}_m$ is the Lagrangian for matter. Using the relations
\begin{equation}
    Q = R + \nabla_{\mu} (Q^{\mu} - \Tilde{Q}^{\mu}),
\end{equation}
and 
\begin{equation}
    \nabla_{\mu} (Q^{\mu} - \Tilde{Q}^{\mu}) \equiv \frac{1}{\sqrt{-g}} \partial_{\mu} \left( \sqrt{-g} \left( Q^{\mu} - \Tilde{Q}^{\mu} \right) \right),
\end{equation}
where $R$ is the Ricci scalar, the STEGR action is dynamically equivalent to GR, except for a boundary term. This boundary term vanishes because the boundary is fixed and the metric variation at the boundary is zero \cite{Capozziello:2022zzh, DAmbrosio:2021zpm}. Importantly, this framework involves second-order field equations, whereas the gravitational field equations in theories with only the Levi-Civita connection are of fourth order \citep{Sotiriou:2008rp}.

Now, we shall consider a well-known extension of STEGR, in which the gravitational Lagrangian $Q$ is replaced by an arbitrary function of non-metricity and the trace of the energy-momentum tensor. Thus, the action integral reads
\begin{equation}\label{eq:action}
    S_{f(Q,T)}=\int \left(\frac{1}{2}\, f(Q,T) +\mathcal{L}_m\right)	\sqrt{-g}\, d^4x.
\end{equation}
Here, $T$ is obtained by contracting the energy-momentum tensor $T_{\mu\nu}$, which is derived by varying $\mathcal{L}_m$ with respect to the metric tensor, i.e.,
\begin{equation}
    T_{\mu \nu} \equiv \frac{-2}{\sqrt{-g}} \frac{\delta(\sqrt{-g} \mathcal{L}_m)}{\delta g^{\mu \nu}}.
\end{equation}

Further, $\delta T = \delta (T_{\mu\nu} g^{\mu\nu}) = (T_{\mu\nu} + \Theta_{\mu\nu}) \delta g^{\mu\nu}$, where
\begin{equation}
    \Theta_{\mu \nu} = g^{\lambda \rho} \frac{\delta T_{\lambda \rho}}{\delta g^{\mu \nu}}.
\end{equation}

Varying the action \eqref{eq:action} with respect to $g_{\rho \mu}$ results in the following field equation

\begin{equation}
\begin{split}
    \frac{-2}{\sqrt{-g}} \nabla_\lambda \left(\sqrt{-g} f_Q P^\lambda_{\;\;\rho \mu}\right) - \frac{1}{2} g_{\rho \mu} f + f_T \left( T_{\rho \mu} + \Theta_{\rho \mu} \right) - f_Q \left( P_{\rho \nu \sigma} Q^{\;\;\nu \sigma}_{\mu} - 2 Q^{\nu \sigma}_{\;\;\rho} P_{\nu \sigma \mu} \right) = T_{\rho \mu},
\end{split}
\end{equation}

where $f_Q = \frac{df}{dQ}$, $f_T = \frac{df}{dT}$, and 

\begin{equation}
\begin{split}
    P_{\lambda \mu \nu} &\equiv \frac{1}{4} \left( -Q_{\lambda \mu \nu} + 2 Q_{\lambda (\mu \nu)} + Q_{\lambda} g_{\mu \nu} - \tilde{Q}_{\lambda} g_{\mu \nu} - \delta_{\lambda (\mu} Q_{\nu)} \right)\\
&= -\frac{1}{2} L_{\lambda \mu \nu} + \frac{1}{4} \left( Q_{\lambda} - \tilde{Q}_{\lambda} \right) g_{\mu \nu} - \frac{1}{4} {\delta^{\lambda}}_{(\mu} Q_{\nu)}
\end{split}
\end{equation}
is the superpotential and the non-metricity scalar $Q$ can be expressed as 
\begin{equation}
\begin{split}
    Q &= -Q^{\alpha \mu \nu} P_{\alpha \mu \nu} \\&= -\frac{1}{4} \left( -Q^{\alpha \nu \rho} Q_{\alpha \nu \rho} + 2 Q^{\alpha \nu \rho} Q_{\rho \alpha \nu} - 2 Q_{\rho} \tilde{Q}^{\rho} + Q_{\rho} Q^{\rho} \right).
\end{split}
\end{equation}

Moreover, varying the gravitational action with respect to the connection yields the field equations

\begin{equation}
    \nabla_\mu \nabla_\nu \left( \sqrt{-g} f_Q P^{\mu \nu}_{\;\;\;\alpha} + 4 \pi H^{\mu \nu}_{\;\;\;\alpha} \right) = 0,
\end{equation}

where 

\begin{equation}
   H_{\mu \nu \lambda} \equiv \frac{\sqrt{-g}}{16 \pi} \left( f_T \frac{\delta T}{\delta \hat{\Gamma}^{\lambda}_{\mu \nu}} + \frac{\delta \left(\sqrt{-g} \mathcal{L}_M\right)}{\delta \hat{\Gamma}^{\lambda}_{\mu \nu}} \right)
\end{equation}

is the hypermomentum tensor. Let us assume that the Universe is isotropic, homogeneous, and spatially flat. The metric that describes this scenario is the Friedmann-Lemaitre-Robertson-Walker (FLRW) metric, which is given by
\begin{equation}\label{eq:metric}
    ds^2 = -N^2(t) dt^2 + a^2(t) \delta_{\mu\nu} dx^\mu dx^\nu.
\end{equation}
Here, $a(t)$ is the scale factor, and $N(t)$ is the lapse function. For this metric, the non-metricity scalar $Q$ takes the forms 
\begin{equation}
    Q = 6\frac{H^2}{N^2}, 
\end{equation}
where $H$ is the Hubble function. In our study, we set $N(t)=1$. Thus, the modified governing equations take the form

\begin{gather}\label{eq:motion1}
3H^2 = \frac{f}{4 f_Q} - \frac{1}{2 f_Q} \left[ \left( 1 + f_T \right) \rho + f_T p \right],\\\label{eq:motion2}
2\dot{H} + 3H^2  = \frac{f}{4f_Q} - \frac{2\dot{f_Q} H}{f_Q} + \frac{1}{2f_Q} \left[ \left( 1 + f_T \right) \rho + \left( 2 + f_T \right) p \right].
\end{gather}

\section{Standard Cosmographic assumptions}\label{sec:III}
It is well known that the cosmographic parameters come from the coefficients of the Taylor series expansion of the scale factor $a(t)$. At the present time $t_0$, it can be defined as \cite{Visser:2004bf,Visser:2015iua}
\begin{equation}\label{eq:a(t)}
    a(t)= 1+ \left.\sum_{n=1}^\infty \frac{1}{n!} \frac{d^na}{dt^n}\right|_{t=t_0} (t-t_0)^n. 
\end{equation}
Considering the leading 4 coefficients, one can find the Hubble, deceleration, jerk, and snap parameters as follows
\begin{itemize}
    \item $H(t)=\frac{a'}{a}$
    \item $q(t)=-\frac{a''}{aH^2}$
    \item $j(t)=\frac{a'''}{aH^3}$
    \item $s(t)=\frac{a''''}{aH^4}.$
\end{itemize}
From now on $(')$ must be understood as the derivative with respect to time $t$. Due to the insufficiency of observational findings, we are not considering the higher-order cosmographic parameters beyond the snap parameter. With the help of the above definitions, one can achieve the derivatives of the Hubble parameter
as 
\begin{gather}
       H'=-H^2 (q+1), \nonumber\\
       H''=H^3 (j+3 q+2) ,\nonumber\\
       H'''=H^4 (-4 j-3 q (q+4)+s-6).
\end{gather}
As the theory involves the two variables $Q$ and $T$, we explicitly find their derivatives, which eventually take the cosmographic form.
\begin{gather}
    Q=6 H^2,\nonumber\\
    Q'=12 H H', \nonumber\\ 
    Q''=12 H H''+12 (H')^2,\nonumber\\ Q'''=12 H H''' +36 H' H''.
\end{gather}

\begin{gather}
    T=\frac{3 {H_0}^2 \Omega_{m0}}{a^3} ,\nonumber\\
    T'=-\frac{9 {H_0}^2 \Omega_{m0} \, a'}{a^4}, \nonumber\\ 
    T''=-\frac{9 {H_0}^2 \Omega_{m0} \, \left(aa''-4{a'}^2\right)}{a^5},\nonumber\\ T'''=-\frac{9 {H_0}^2 \Omega_{m0} \, \left(20{a'}^3-12aa'a''+a^2a'''\right)}{a^6}.
\end{gather}
Finally, we define the time derivatives of the scale factor $(a(t))$ as follows
\begin{gather}
    a' = aH,\nonumber\\
    a'' = H a' + a' H',\nonumber\\
a''' = 2 a' H' + H a'' + a H'',\nonumber\\
a''''=3 a''H' + 3 a' H'' + H a''' + a H'''.
\end{gather}
The quantities we obtained in this section will play a vital role in the cosmographic construction of the $f(Q, T)$ theory in the upcoming sections.

\section{Cosmography with Chebyshev Polynomials}\label{sec:IV}

Though in cosmography using Taylor series approximation is the most adopted approach,
the restriction $z<1$ on the convergence makes dealing with the data with higher redshifts complicated. This is why we consider the Chebyshev polynomial, which has already proven its suitability at higher redshifts compared to the Taylor series and Pad\'{e} approximations \cite{Capozziello:2017nbu}.

\subsection{Construction of the two variables functional using the first kind Chebyshev Polynomial}

The Chebyshev series converges remarkably fast, exponential convergence for analytic functions while providing a global approximation over an interval, making it particularly effective for capturing the overall behavior of a function compared to the Taylor series. The Chebyshev polynomials (first kind) are defined as 
\begin{equation}
    \mathcal{T}_{n}(x)=cos(n\theta),
\end{equation}
where $\theta=cos^{-1}(x)$. One of the most convenient properties of the first kind 
Chebyshev polynomials is orthogonality in the domain $[-1,1]$ with respect to the inner product
\begin{equation}
    <f_1,f_2>=\int_{-1}^1 f_1(x) f_2(x)\frac{dx}{(1-x^2)^\frac{1}{2}},
\end{equation}
where $1/(1-x^2)^{1/2}$ is the weight function. To find the polynomials, one can use the following recurrence relation 
\begin{gather}
    \mathcal{T}_0(x)=1,\nonumber\\
    \mathcal{T}_1(x)=x,\nonumber\\
    \mathcal{T}_{n+1}(x)=2x\mathcal{T}_n(x)-\mathcal{T}_{n-1}(x).
    \label{eq:rec}
\end{gather}
Using the above recurrence relation, we explicitly mention the remaining polynomials up to $\mathcal{T}_4(x)$ below, which will be further used to construct the functional form and luminosity distance in terms of the Chebyshev series.
\begin{gather}
    \mathcal{T}_2(x)=2x^2-1,\nonumber\\
    \mathcal{T}_3(x)=4x^3-3x,\nonumber\\
    \mathcal{T}_4(x)=8x^4-8x^2+1.
    \label{eq:reccont}
\end{gather}

To deal with the functions of two variables, we start with the definition of the Chebyshev series for the function $f(Q, T)$ as follows \cite{Ernest:2015}
\begin{equation}\label{eq:fcheb}
    f(Q,T)\sim \sum_{i,j=0}^\infty \alpha_{i,j} \,\ \mathcal{T}_i(Q) \,\ \mathcal{T}_j(T).
\end{equation}

The coefficients of the above expression can be calculated using the following formulae

\begin{align}
\label{eq:chebcoeff}
    \alpha_{0,0}&=\frac{1}{\pi^2}\int_{-1}^1 \int_{-1}^1 g(Q,T) *  w(Q,T) \,\ dQ \,\ dT \nonumber\\
    \alpha_{0,j}&=\frac{2}{\pi^2}\int_{-1}^1 \int_{-1}^1 g(Q,T) *  w(Q,T)*\mathcal{T}_{j}(T)\,\ dQ \,\ dT\nonumber\\
    \alpha_{i,0}&=\frac{2}{\pi^2}\int_{-1}^1 \int_{-1}^1 g(Q,T) * w(Q,T)*\mathcal{T}_{i}(Q)\,\ dQ \,\ dT\nonumber\\
    \alpha_{i,j}&=\frac{4}{\pi^2}\int_{-1}^1 \int_{-1}^1 g(Q,T) *  w(Q,T)*\mathcal{T}_{i}(Q)*\mathcal{T}_{j}(T)\,\ dQ \,\ dT  
\end{align}

where $g(Q,T)$ is the Taylor series expansion of the function $f(Q,T)$ and $w(Q,T)$ is the weight function, defined as $w(Q,T)=\left((1-Q^2)^{1/2}(1-T^2)^{1/2}\right)^{-1}$. Further to reduce complexity, a minimally coupled form $\gamma(Q)+\eta(T)$ of the theory is assumed. Now the Taylor series approximation up to $4^{th}$ order can be expressed as 

\begin{multline}
\label{eq:ftaylor}
    g(Q,T)\sim \gamma+\frac{1}{2} {\gamma}^{(1)} (Q-Q_0)+\frac{1}{6} {\gamma}^{(2)} (Q-Q_0)^2+\frac{1}{24} {\gamma}^{(3)} (Q-Q_0)^3+\frac{1}{120} {\gamma}^{(4)} (Q-Q_0)^4+\\\eta+\frac{1}{2} {\eta}^{(1)} (T-T_0)+\frac{1}{6} {\eta}^{(2)} (T-T_0)^2+\frac{1}{24} {\eta}^{(3)} (T-T_0)^3+\frac{1}{120} {\eta}^{(4)} (T-T_0)^4.
\end{multline}
Here $\gamma^{(n)}$ and $\eta^{(n)}$ represents the $n^{th}$ order derivatives of $\gamma$ and $\eta$, respectively.
By incorporating \eqref{eq:chebcoeff} and \eqref{eq:ftaylor} in the Chebyshev series \eqref{eq:fcheb}, one can finally achieve
\begin{multline}
    f(Q,T) \sim \gamma+\frac{1}{120} (Q-Q_0) \left(60 {\gamma}^{(1)}+(Q-Q_0)\times \right.
    \left. \left(20 {\gamma}^{(2)}+(Q-Q_0) \left(5 {\gamma}^{(3)}+{\gamma}^{(4)} Q-{\gamma}^{(4)} Q_0\right)\right)\right)+\\\eta+\frac{1}{120} (T-T_0) \left(60 {\eta}^{(1)}+(T-T_0)\times \right.  \left. \left(20 {\eta}^{(2)}+(T-T_0) \left(5 {\eta}^{(3)}+{\eta}^{(4)} T-{\eta}^{(4)} T_0\right)\right)\right).
\end{multline}

\subsection{Construction of the Luminosity distance using the first kind Chebyshev Polynomial}
A relation between powers of $z$ and the Chebyshev polynomials can be expressed as \cite{Litinov:1994,Capozziello:2017nbu}
\begin{equation}
  z^n=2^{1-n}{\sum_{\underset  {k\,\ \equiv \,\ n(mod 2)}{k=0}}^{n\;\ '}} \binom{n}{\frac{n-k}{2}} \mathcal{T}_{k}(z),
\end{equation}
here the prime represents that if the contribution of $k=0$ appears, it needs to be halved. Before moving to the Chebyshev series we define the Taylor series expansion of the luminosity distance in terms of cosmographic parameters as
\begin{multline}\label{eq:dL}
    d_L(z)=\frac{cz}{H_0} \left(1+\frac{1}{2!} (1-q_0) z-\right.\left. \frac{1}{3!} z^2 \left(j_0-3 {q_0}^2-q_0+1\right)+\right.\\\left. \frac{1}{4!} z^3 \left(10 j_0 q_0+5 j_0-15 {q_0}^3-15 {q_0}^2-2 q_0+s_0+2\right)+\right.\left.\mathcal{O}(z^4) \right).
\end{multline}

Finally, using the above ingredients along with the first five terms of the recurrence relation \eqref{eq:rec}, one can obtain the luminosity distance for the Chebyshev polynomials as

\begin{equation}
    d_L(z)=\frac{c}{H_0}\sum_{n=0}^4 c_n \mathcal{T}_n(z),
\end{equation}
where 
\begin{gather}
\begin{aligned}
 c_0&=\alpha(54+15j_0(1+2q_0)-9q_0(6+5q_0(1+q_0))+3s_0),\\
    c_1&=\alpha(168-24j_0+24q_0+72q_0^2),\\
    c_2&=\alpha(56+20j_0(1+2q_0)-4q_0(14+15q_0(1+q_0))+4s_0),\\
    c_3&=\alpha(-8-8j_0+8q_0(1+3q_0)),\\
    c_4&=\alpha(2+5j_0(1+2q_0)-q_0(2+15q_0(1+q_0))+s_0),
\end{aligned}
\end{gather}
and $\alpha=\frac{1}{192}$.
We intend to constrain the $f(Q,T)$ theory by using the above 4th order $d_L(z)$ expression in terms of the cosmographic parameters. 

\subsection{Cosmographic parameters}
We start this section by incorporating the assumed minimally coupled form in the motion equations \eqref{eq:motion1} \& \eqref{eq:motion2}. Hence the revised motion equations at the present time reads
\begin{gather}\label{eq:revmotion1}
\begin{split}
    \gamma+\eta= 6 {H_0}^2 (2\gamma^{(1)}+\Omega_{m0})+\frac{1}{1+\eta^{(1)}} \times\left[4 H_0 \eta^{(1)} \left(\gamma^{(2)}-H_0(1+q_0) \gamma^{(1)}\right)\right],
\end{split}\\\label{eq:revmotion2}
   {H_0}^2(2-q_0)=\frac{-4 H_0 \gamma^{(2)}+\gamma+\eta+6{H_0}^2 \Omega{m0}}{4 \gamma^{(1)}}.
\end{gather}
By solving the above equations and their further derivatives, one can obtain the cosmographic parameters in terms of the unknowns as follows\\
\
\\
\text{\textbf{Deceleration:}}
\begin{equation}
\begin{split}\label{eq:deceleration}
    q_0=&\left(1/\left(48 \gamma^{(1)} {H_0}^3 (1+\eta^{(1)})\right)\right)\left(-6 \gamma^{(1)} (1+8 {H_0}^3) (1+\eta^{(1)}) \right. \left.+24 \gamma^{(1)} {H_0}^2 \eta^{(2)} {\Omega_{m0}}+  {\Omega_{m0}} \left(-72 {H_0}^3 (1+\eta^{(1)})^2+ \right.\right.\\&\left.\left.(1+\eta^{(1)}) (2+\eta^{(1)})+6 {H_0}^2 (1+2 \eta^{(1)}) \eta^{(2)} {\Omega_{m0}}\right)+ \right. \left. \frac{1}{2} \left(-576 \gamma^{(1)} {H_0}^3 (1+\eta^{(1)}) \left(4 \gamma^{(1)} \left((-1+4 {H_0}^3) (1+\eta^{(1)})- \right.\right.\right.\right.\\&\left.\left.\left.\left.6 {H_0}^2 \eta^{(2)} {\Omega_{m0}}\right)+ {\Omega_{m0}} \left(48 {H_0}^3 (1+\eta^{(1)})^2-(1+\eta^{(1)}) (2+3 \eta^{(1)}) \right.\right.\right.\right. \left.\left.\left.\left.-6 {H_0}^2 (1+2 \eta^{(1)}) \eta^{(2)} {\Omega_{m0}}\right)\right)+\right.\right.\\ &\left.\left.  4 \left(-6 \gamma^{(1)} (1+8 {H_0}^3) (1+\eta^{(1)}) \right.\right.\right.\left.\left.\left.+24 \gamma^{(1)} {H_0}^2 \eta^{(2)} {\Omega_{m0}}+   {\Omega_{m0}} \left(-72 {H_0}^3 (1+\eta^{(1)})^2+ \right.\right.\right.\right.\\&\left.\left.\left.\left.(1+\eta^{(1)}) (2+\eta^{(1)})+6 {H_0}^2 (1+2 \eta^{(1)}) \eta^{(2)} {\Omega_{m0}}\right)\right)^2\right)^{1/2}\right),
\end{split}
\end{equation}

\text{\textbf{Jerk:}}
\begin{equation}\label{eq:jerk}
\begin{split}
    j_0=&2-12 {H_0}^3 (-2+q_0) (1+q_0)^2+q_0 (2+q_0)+ \left(48 \gamma^{(3)} H_0 (1+q_0)+  3 {\Omega_{m0}} \left(-2+\eta^{(1)}+2 \eta^{(1)} q_0-\right.\right.\\&\left.\left.48 {H_0}^3 (1+\eta^{(1)}) (-2+q_0) (1+q_0)\right)\right)\left(1/4 \gamma^{(1)}\right), 
\end{split}
\end{equation}
\text{\textbf{Snap:}}
\begin{equation}\label{eq:snap}
\begin{split}
    s_0=&-5-8 q_0-288 {H_0}^6 (-2+q_0) (1+q_0)^4-q_0 (j_0+3 q_0)+ (1/{\gamma^{(1)}}^2)\left(54 {H_0}^3 (1+\eta^{(1)}) {\Omega_{m0}}\right.\\&\left.\times (1+q_0) \left(16 \gamma^{(3)} H_0 (1+q_0)+  {\Omega_{m0}} \left(-2+\eta^{(1)}+\right.\right.\right.\left.\left.\left.2 \eta^{(1)} q_0-48 {H_0}^3 (1+\eta^{(1)}) (-2+q_0) (1+q_0)\right)\right)\right)  \\& +12 {H_0}^3 (1+q_0) \left(-7-2j_0+q_0 j_0-7q_0 +6 {q_0}^2+3{q_0}^3\right) +(3/4 \gamma^{(1)})\left(-192 \gamma^{(4)} {H_0}^3 (1+q_0)^2+3 {H_0}^2 \eta^{(2)} {\Omega_{m0}}^2 \right.\\&\left. \times(3+2 q_0)^2-  16 \gamma^{(3)} H_0 \left(5+j_0+3 q_0 \left(3+q_0-\right.\right.\right.\left.\left.\left.12 {H_0}^3 (1+q_0)^2\right)\right)+   {\Omega_{m0}} \left(16-2 \eta^{(1)} (-2+j_0)+\right.\right.\\&\left.\left.3 \left(6+\eta^{(1)}\right) q_0+2 \left(2+\eta^{(1)}\right) {q_0}^2-   2304 {H_0}^6 \left(1+\eta^{(1)}\right)\right.\right.\left.\left. \times(-2+q_0) (1+q_0)^3+24 {H_0}^3 \left(-2 (8+2 j_0+9 q_0)+  \right.\right.\right.\\&\left.\left.\left. 2 q_0 \left(j_0+q_0 (5+3 q_0)\right)+\eta^{(1)} \left(-13+2 j_0 (-2+q_0)\right.\right.\right.\right.\left.\left.\left.\left.+q_0 \left(-10+q_0 \left(17+8 q_0\right)\right)\right)\right)\right)\right).
\end{split}
\end{equation}

\section{Observational constraints using Cosmography}\label{sec:V}
In the previous section, we constrained the free variables of the constructed functional form using the cosmographic parameters. Though one can use the present values of the cosmographic parameters from observations to get a system of linear equations, it will lead to an infinite number of solutions. So to achieve a definitive range for the unknowns, we find the Chebyshev series for luminosity distance. By using the luminosity distance, we intend to perform the MCMC analysis for the Pantheon+SH0ES data which may lead us to the best-fit ranges for the appeared unknowns.

Observational data are vital in assessing the nature and dynamics of the Universe, and we have seen rapid advancements in recent decades. With increasing sensitivity in experimental data from various astronomical and cosmological surveys, the core of modern cosmology is being refined. This progress allows us to achieve more precise measurements of cosmological parameters. Developments in observational cosmology have opened new pathways, providing a provision for identifying a cosmological theory that aligns with observational evidence.

\subsection{PANTHEON+SH0ES}
The Pantheon+SH0ES dataset, as detailed in references \cite{Riess:2021jrx,Malekjani:2023dky,Brout:2022vxf,Brout:2021mpj,Scolnic:2021amr}, includes distance moduli derived from 1701 light curves of 1550 Type Ia supernovae (SNeIa) collected across 18 different surveys. These light curves cover a redshift range from $0.001 \leq z \leq 2.2613$. Significantly, the dataset contains 77 light curves from galaxies that also host Cepheids. One of the key advantages of the Pantheon+SH0ES dataset is its effectiveness in constraining the Hubble constant $H_0$ along with other free parameters. We estimate the theoretical distance modulus for the SNeIa sample using the equation

\begin{equation}\label{eq:mu}
    \mu(z) = 5\log_{10}\left(\frac{d_L(z)}{1\ \text{Mpc}} \right) + 25,
\end{equation} 

where $d_L$ is defined as in \eqref{eq:dL}.
The free parameters $(H_0, \gamma^{(1)}, \gamma^{(3)}, \gamma^{(4)}, \eta^{(1)}, \eta^{(2)} \Omega_{m0})$ are constrained by applying equations \eqref{eq:deceleration}-\eqref{eq:snap} to \eqref{eq:mu}. As a result, the distance residual $\Delta\mu$ is expressed as 

 \begin{equation}
        \Delta\mu_i=\mu_i-\mu_{th}(z_i).
    \end{equation}

When analyzing data from the SNeIa sample, a degeneracy is observed between the parameters $H_0$ and $M$. To address this issue, a modification is applied to the SNeIa distance residuals \cite{Perivolaropoulos:2023iqj,Brout:2022vxf} as follows

\begin{equation}
    \Delta\Tilde{\mu} = \begin{cases} 
        \mu_i - \mu_i^{\text{Ceph}}, & \text{if } i \text{ is a Cepheid host} \\
        \mu_i - \mu_{\text{th}}(z_i), & \text{otherwise}
    \end{cases}
\end{equation}

Here, $\Delta\Tilde{\mu}$ denotes the modified distance residual, and $\mu_i^{\text{Ceph}}$ represents the distance modulus of the Cepheid host for the $i^{\text{th}}$ SNeIa. The $\chi^2$ function for SNeIa is then given by

\begin{equation}\label{Eq:ChiSN}
    \chi^2_{\text{SNeIa}} = \Delta\mu^T (C_{\text{stat+sys}}^{-1})\Delta\mu.
\end{equation}

\subsection{Results}
One can notice that the parameters $\gamma$, $\eta$, $\gamma^{(2)}$, $\eta^{(3)}$, and $\eta^{(4)}$ are missing from the MCMC analysis. However, it is to be noted that the parameters develop a dependency on the remaining parameters when we do further derivatives of the motion equations which is why it is not necessary to consider them. For instance, one can find $\gamma^{(2)}$ by solving \eqref{eq:revmotion1} and \eqref{eq:revmotion2}. In a similar fashion, the dependent parameter ranges can be obtained conveniently.

\begin{figure}
    \centering
    \includegraphics[width=\textwidth]{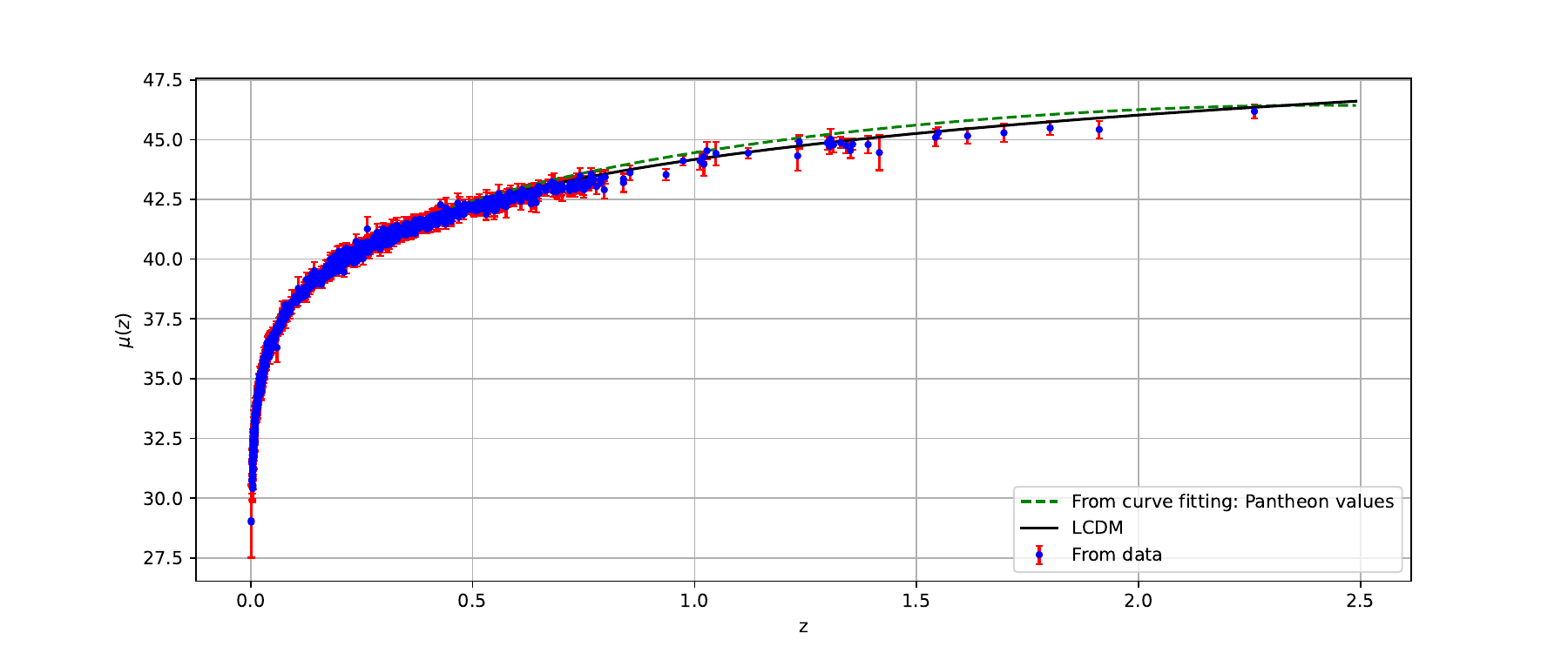}
    \caption{Curve fitting of the distance modulus function against the 1701 data points of PANTHEON+SH0ES and $\Lambda$CDM model.}
    \label{fig:curve}
\end{figure}

\begin{figure}
    \centering
    \includegraphics[width=\textwidth]{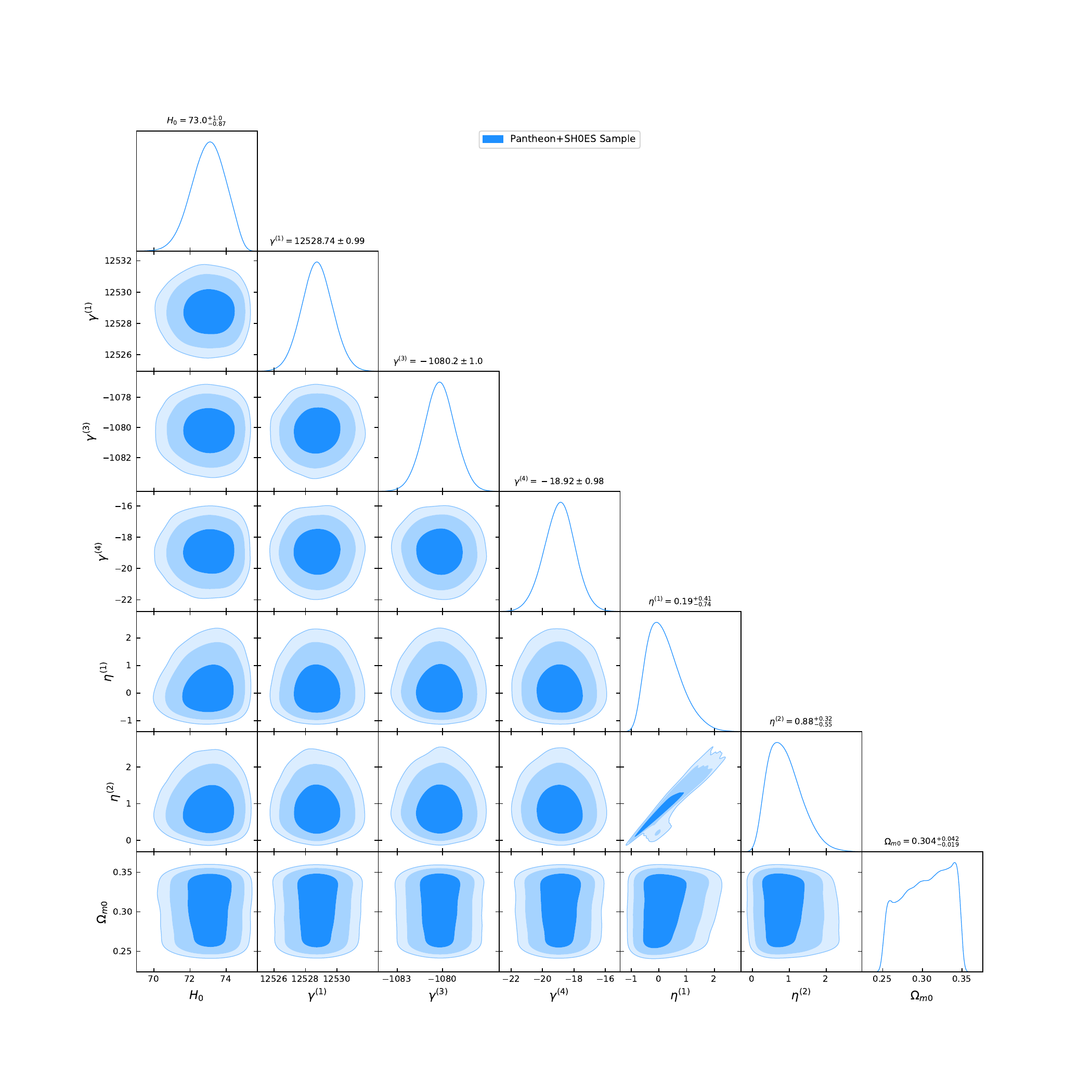}
    \caption{Upto $3-\sigma$ CL contours and posterior distributions from the MCMC analysis of PANTHEON+SH0ES sample.}
    \label{fig:contour}
\end{figure}

The contours up to $3-\sigma$ CL in \autoref{fig:contour} are obtained from the MCMC technique by using the chi-square function of PANTHEON+SH0ES data. The innermost dark shaded region is $1-\sigma \, (68\%)$ CL, the middle region is $2-\sigma \, (95\%)$ CL and the outer light shaded region is $3-\sigma \, (99.7\%)$ CL.  We summarize the best fit $1-\sigma$ confidence level ranges for the parameters in \autoref{table1}. Moreover, we depict from \autoref{fig:curve} that the distance modulus function for our constrained theory perfectly aligns with the 1701 points of PANTHEON+SH0ES sample and the standard $\Lambda$CDM model.

Further, we have performed a statistical comparison of our model to the standard model by using the Bayesian techniques Akaike Information Criterion (AIC) and Bayesian Information Criterion (BIC). The tools are defined as $AIC= \chi^2_{min}+2d$ and $BIC= \chi^2_{min}+ d ln N,$ where $d$ is the number of free parameters, and $N$ is the sample size of the corresponding dataset. To get the evidence how strongly the models are supported by data, one has to find the difference $\Delta AIC=|{{AIC}_{\Lambda CDM}-{AIC}_{MODEL}}|$ and $\Delta BIC=|{{BIC}_{\Lambda CDM}-{BIC}_{MODEL}}|$ (See \cite{Kolhatkar:2024oyy} for the acceptable ranges for evidence). For our results, we achieved $\Delta AIC=1.88$, which indicates strong evidence in favor of the model. Since our model contains a high number of parameters as compared to the standard model, the $\Delta BIC$ value appears as $14.43$, which is slightly higher and does not provide any supportive evidence. 
    
\begin{table}
 \centering
 \caption{ Best fit range of the parameters with $1-\sigma$ confidence level.}
 
 \label{table1}
    \begin{tabular}{|c||c|}
    \hline
    
          & $Pantheon+SH0ES$ \\
    \hline
    \hline

    $H_0$ & $(72.13,74)$ \\
    \hline
    
     $\gamma^{(1)}$ & $(12527.75,12529.73)$  \\
    \hline

     $\gamma^{(3)}$ & $(-1081.2,-1079.2)$ \\
    \hline
    
     $\gamma^{(4)}$ & $(-19.9,-17.94)$ \\
    \hline
    
     $\eta^{(1)}$ & $(-0.55,0.6)$  \\
    \hline
    
     $\eta^{(2)}$ & $(0.33,1.2)$  \\
    \hline
    
     $\Omega_{m0}$ & $(0.285,0.346)$ \\
    \hline
   
    \end{tabular}
\end{table}

\section{Conclusion}\label{sec:VI}
In recent times, many successful modified theory models have proved their efficacy in describing gravity. For example, models by Starobinsky \cite{Starobinsky:1980te} and Linder \cite{Linder:2016wqw} have successfully described various physical phenomena. To uncover the fundamental principles governing the Universe, reconstructing gravitational theory offers a promising approach. This reconstruction can be accomplished by either constraining the model parameters of pre-assumed functional forms or employing the Raychaudhuri equations, both of which are dependent on specific gravity models.

However, it is crucial to adopt a more general perspective that does not rely on particular models. There are several methods for exploring the dynamics within modified theories without assuming a specific functional form. These methods include purely statistical techniques such as Gaussian processes and neural networks, as well as cosmographic approaches. By utilizing cosmographic parameters like the Hubble parameter, deceleration, jerk, and snap, cosmographic techniques are very helpful in developing a theory with a model-independent approach. We have used these techniques to constrain the extension of symmetric teleparallel theory in this work. The main significance is the utilization of the Chebyshev series for two variables to reconstruct the functional form of $f(Q, T)$. The Chebyshev series provides exceptionally fast, exponential convergence for analytic functions and provides a global approximation over an interval, making it ideal for capturing the overall behavior of a function compared to the Taylor series. It is less sensitive to singularities near the domain and exhibits robust numerical stability due to the orthogonality of Chebyshev polynomials. This method is particularly well-suited for efficiently approximating smooth functions across a wide range. On the other hand, Pad\'e polynomials also perform well in certain contexts, as supported by existing references \cite{Capozziello:2020ctn, Hu:2024qnx}. However, our work is distinguished from prior cosmography literature by addressing functions with two variables. Researchers have predominantly explored Pad\'e, Taylor, and Chebyshev series cosmography for single-variable functions. Here, we introduce a novel methodology that applies the Chebyshev series to functions of two variables. For such multivariable functions, the Pad\'e expansion becomes more complex, and parameter estimation through statistical processes becomes increasingly infeasible, highlighting the advantages of our approach.

Further from the motion equations, we found a set of solutions for the cosmographic parameters involving the free variables that appeared in our reconstructed form. We obtained the luminosity distance in terms of the present-time cosmographic parameters $(H_0,\, q_0,\, j_0,\, s_0)$. To constrain our theory, we conducted a Markov Chain Monte Carlo analysis because it represents a class of probability distribution sampling techniques based on the construction of Markov chains. The distance modulus function $(\mu(z))$, which can be obtained from $d_L(z)$ is used to minimize the chi-square function and to compare with the $1701$ points of PANTHEON+SH0ES sample. These data are some of the most precise tools available for measuring key cosmological parameters. However, they are not immune to systematic errors, and careful attention is required to identify, quantify, and mitigate these uncertainties. Some of these errors are intrinsic to the methods used, while others arise from the inherent complexities of measuring cosmic distances. Finally, the AIC test provided strong evidence in favor of our model, which indicates the alignment of the model to the data. We found that the results from MCMC analysis make an excellent match to the $\Lambda$CDM model and the data points.

\section*{Data availability} There are no new data associated with this article.

\acknowledgments  SSM acknowledges the Council of Scientific and Industrial Research (CSIR), Govt. of India for awarding Junior Research fellowship (E-Certificate No.: JUN21C05815). PKS acknowledges Anusandhan National Research Foundation (ANRF), Department of Science and Technology, Government of India for financial support to carry out Research project No.: CRG/2022/001847. NSK, and VV acknowledge DST, New Delhi, India, for its financial support for research facilities under DST-FIST-2019. We are very much grateful to the honorable referees and to the editor for the illuminating suggestions that have significantly improved our work in terms
of research quality, and presentation. 
\nocite{}
\bibliography{main}

\end{document}